\documentclass[preprint,aps,12pt,preprintnumbers,eqsecnum,nofootinbib]{revtex4}
\usepackage{graphicx}
\DeclareGraphicsExtensions{.pdf,.eps}
\usepackage{subfigure}

\usepackage{color}
\usepackage{amssymb,amsmath}
\usepackage{epstopdf}

\newcommand{\beq}{\begin{equation}}
\newcommand{\enq}{\end{equation}}

\begin{document}
%
\title{\vspace*{0.5in} 
Multi-field inflation  and the field-space metric
\vskip 0.1in}
\author{Joshua Erlich}\email[]{jxerli@wm.edu}
\author{Jackson Olsen}\email[]{jcolsen@email.wm.edu}
\author{Zhen Wang}\email[]{zwang01@email.wm.edu}

\affiliation{High Energy Theory Group, Department of Physics,
College of William and Mary, Williamsburg, VA 23187-8795}
\date{September dd, 2015}
\date{\today}

\begin{abstract}
Multi-field inflation models include a variety of scenarios for how  inflation proceeds and ends. Models with the same potential but different  kinetic terms are common in the literature.  We compare  spiral inflation and Dante's inferno-type models, which differ only in their field-space metric.  We justify a single-field effective description in these models and relate the single-field  description to a mass-matrix formalism. We note the effects of the  nontrivial field-space metric on inflationary observables, and  consequently on the viability of these models. We also note a duality between spiral inflation and Dante's inferno models with different potentials.

\end{abstract}
\maketitle

\section{Introduction}
Predictions for inflationary observables depend on both the field-space metric and potential of the fields responsible for the inflationary dynamics. Nontrivial kinetic terms  which modify the field-space metric  arise in many ways: from radiative corrections, from a higher-dimensional origin of the fields, or simply from a  field redefinition. Supersymmetric models of inflation typically include nontrivial Kahler potentials which modify the field-space metric, as in  Ref.~\cite{Li:2013moa} and many of the models reviewed in Refs.~\cite{Yamaguchi:2011kg}. A covariant approach to analyzing fluctuations in an inflationary setting with nontrivial kinetic terms was developed in Ref.~\cite{Gong:2011uw}.
Here we compare two  classes of multi-field inflation models which differ only in their kinetic terms, and we discuss some of the lessons learned from these examples.   
We  justify a single-field effective description of these models and derive a mass matrix appropriate for calculation of inflationary observables in these models.

Although the observation by the BICEP2 collaboration of B-modes in the polarization of microwave radiation~\cite{Ade:2014xna} can be attributed to scattering off of galactic dust~\cite{Flauger:2014qra,Mortonson:2014bja} as demonstrated by the Planck experiment~\cite{Adam:2014oea}, current and proposed experiments such as PIPER~\cite{Lazear:2014bga} remain sensitive to signatures of primordial gravitational waves produced during inflation. In  slow-roll inflation models, the Lyth bound~\cite{Lyth:1996im} implies that the inflaton field typically varies over super-Planckian values if  sufficiently large power in gravitational waves is produced during inflation. This makes it difficult to describe such an inflationary scenario in terms of an effective field theory valid below the Planck scale. There are several ways to evade the Lyth bound, for example if the slow-roll parameter $\epsilon$ increases for some period during inflation, as happens in certain hybrid inflation models~\cite{Carrillo-Gonzalez:2014tia,Carone:2014lba}, or if the inflaton is embedded in a multi-field model in which one of the fields has a discrete shift symmetry, as in axion-monodromy models~\cite{McAllister:2008hb}. Simplified models of the latter type were developed in Refs.~\cite{Kim:2004rp,Berg:2009tg}.

\begin{figure}[t]
\includegraphics[width = 0.5\textwidth]{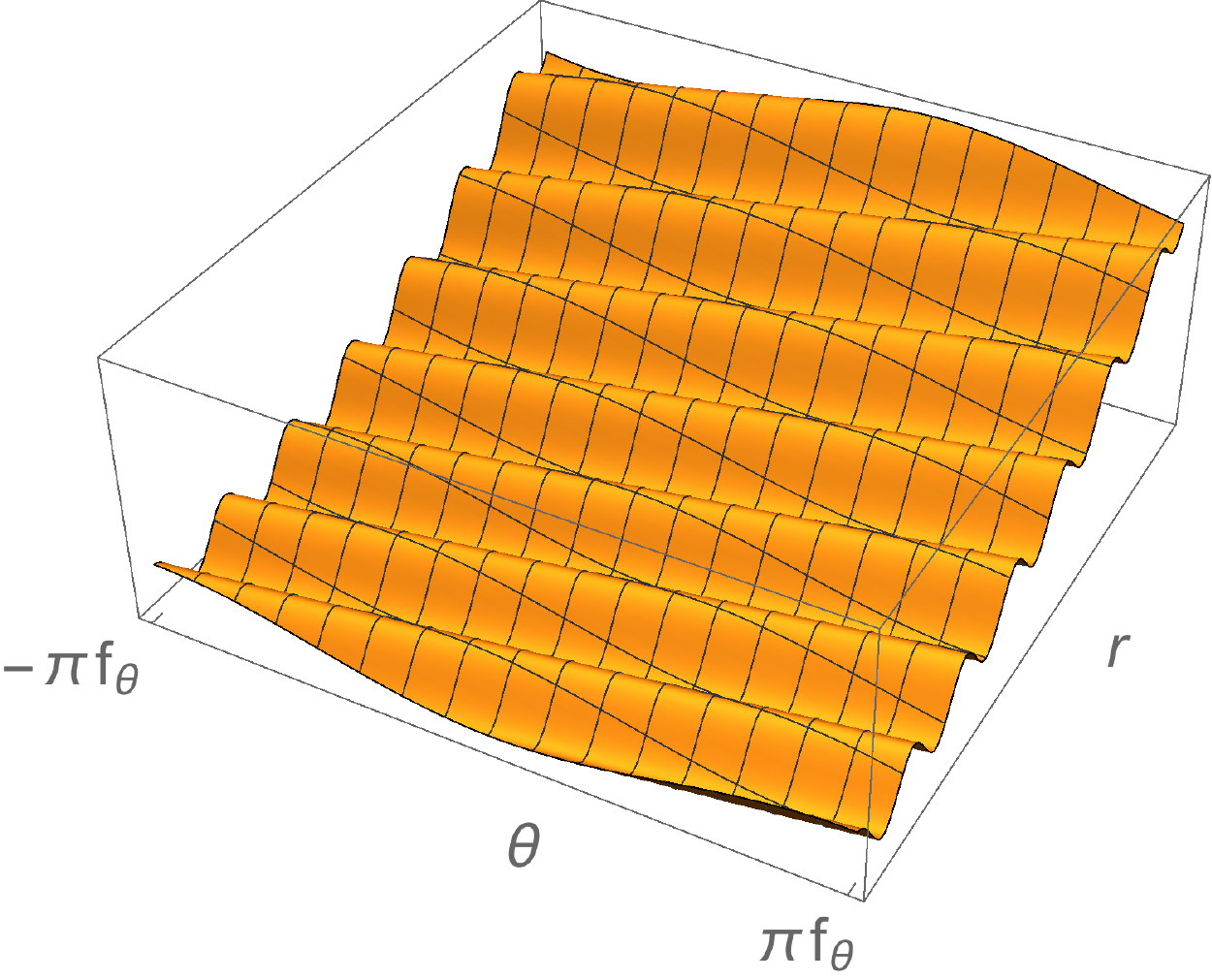}
\caption{The potential as a function of $r$ and $\theta$ in Dante's Inferno with a quadratic shift-symmetry-breaking potential $W(r)=\frac{1}{2}m^2r^2$, as in Ref.~\cite{Berg:2009tg}.}
\label{fig:DI_potential}
\end{figure}

Inflationary models  based on one or more pseudo-Nambu-Goldstone bosons have a long history (for example, Refs.~\cite{Freese:1990rb,Adams:1992bn,Kaplan:2003aj,ArkaniHamed:2003mz,Dimopoulos:2005ac,Silverstein:2008sg}). The Dante's inferno model, developed in Ref.~\cite{Berg:2009tg}, includes two axion fields which evolve along a trench in the potential during inflation, as in Fig.~\ref{fig:DI_potential}. 
The two axions $r$ and $\theta$ in Dante's inferno have canonical kinetic terms, \begin{equation}
{\cal L}_{{\rm DI}}=\frac{1}{2}(\partial_\mu r)^2+\frac{1}{2}(\partial_\mu \theta)^2-V(r,\theta). \label{eq:L_DI}
\end{equation}
The potential has the form \begin{equation}
V(r,\theta)=W(r)+\Lambda^4\left[1-\cos\left(\frac{r}{f_r}-\frac{\theta}{f_\theta}\right) \right]\label{eq:DI-potential},\end{equation}
where the discrete shift symmetry of the axion field $r$ is broken by the term $W(r)$ in the potential. A string-theoretic scenario which gives rise to the Dante's Inferno model was presented in Ref.~\cite{Berg:2009tg}, in which the shift-symmetry-breaking potential $W(r)$ describes the axion on an NS5 brane wrapped on a 2-cycle belonging to a family of homologous 2-cycles which extend into a warped throat geometry. 

\begin{figure}[t]
\includegraphics[width = 0.5\textwidth]{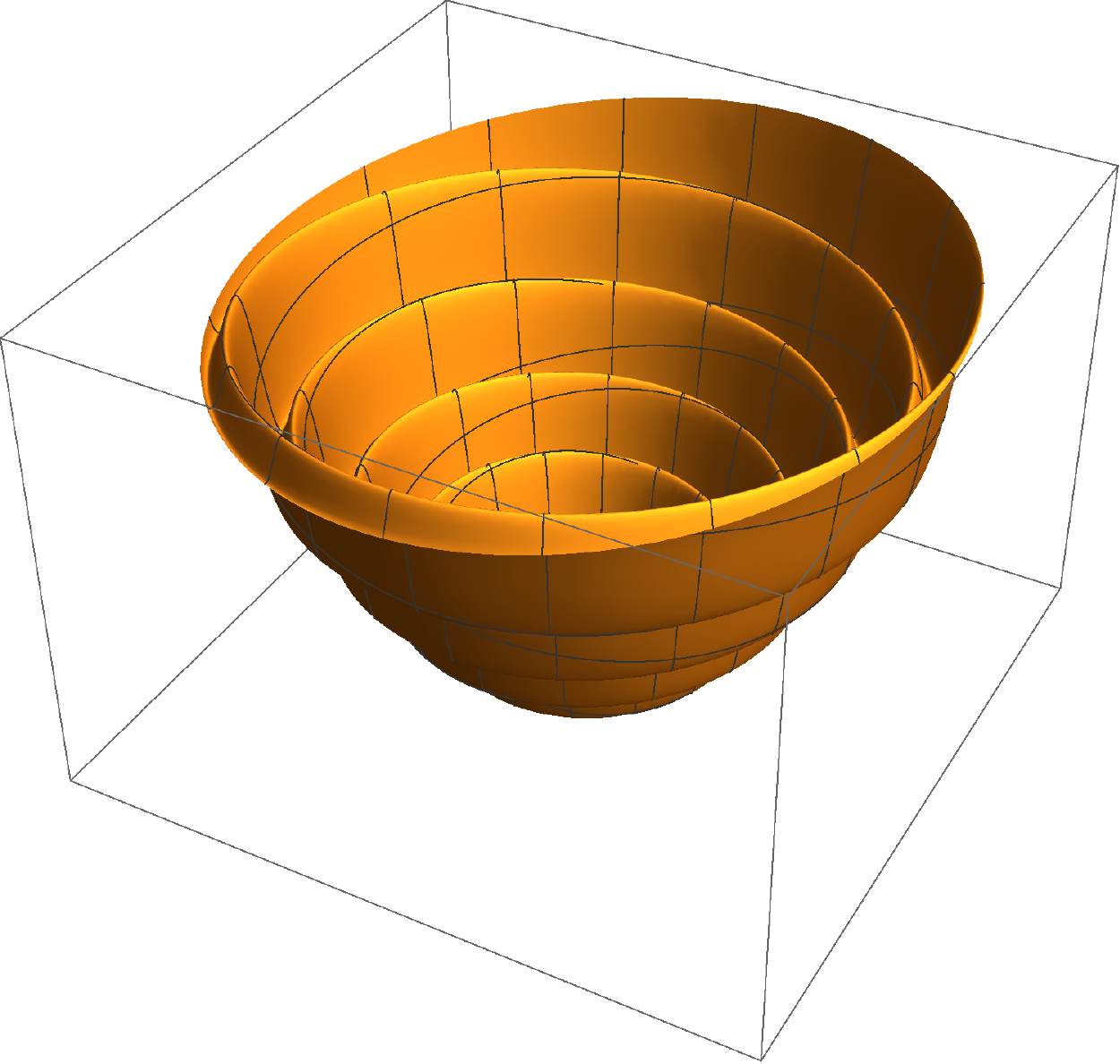}
\caption{The potential as a function of $r$ and $\theta$ in a spiral inflation model with a quadratic shift-symmetry-breaking potential $W(r)=\frac{1}{2}m^2r^2$. The fields $r$ and $\theta$ are represented in polar coordinates.}
\label{fig:SI_pot}
\end{figure}

We will consider a generalization of the potential Eq.~(\ref{eq:DI-potential}) of the form, \begin{equation}
V(r,\theta)=W(r)+\Lambda^4\left[1-\cos\left(\frac{r^n}{f_r^n}-\frac{\theta}{f_\theta}\right) \right]\label{eq:potential}.\end{equation} This class of potentials  appears in models with a complex scalar field and a single anomalous U(1) symmetry, as in the axion inflation model of Ref.~\cite{McDonald:2014nqa}. In this case, the real fields $r/\sqrt{2}$ and $\theta$ are the magnitude and phase, respectively, of a canonically normalized complex scalar field $\Phi=re^{i\theta}/\sqrt{2}$, in which case we  take $f_\theta=1$.  The trench spirals around the potential as in Fig.~\ref{fig:SI_pot}.  
The kinetic terms for the real scalars in these spiral inflation models are non-canonical, taking the form \begin{equation}
{\cal L}_{{\rm SI}}=|\partial_\mu\Phi|^2-V(\Phi)=\frac{1}{2}(\partial_\mu r)^2+\frac{1}{2}r^2 (\partial_\mu \theta)^2 -V(r,\theta).
\label{eq:L_SI}\end{equation}
The additional factor of $r^2$ in the kinetic term for $\theta$ can have important consequences, even affecting the phenomenological viability of these models, as we will see. In this paper we compare the predictions for a number of two-field models with canonical and non-canonical kinetic terms of the form Eq.~(\ref{eq:L_DI}) and Eq.~(\ref{eq:L_SI}).  These include models which are effectively either chaotic inflation or hybrid inflation models. Hybrid inflation models of this type include  Dante's waterfall \cite{Carone:2014cta}  and certain spiral inflation \cite{Barenboim:2014vea,Barenboim:2015zka,Barenboim:2015lla} models. In the case of spiral inflation we will take $f_\theta=1$ so that the potential is periodic in $\theta\rightarrow\theta+2\pi$, while there is a monodromy in shifts of $r$. The qualitative difference between these models can be described in terms of the trajectories of the fields which evolve during inflation:  In the Dante's inferno and Dante's waterfall scenarios the fields evolve along an approximately linear trajectory in the canonically normalized field space, whereas in spiral inflation models the fields evolve along a nearly circular trajectory. 
In a single-field effective description these are chaotic inflation models, but one must take care in the analysis of models with changing inflaton direction as in spiral inflation.
 
In Sec.~\ref{sec:1-field} we describe the single-field effective description of these multi-field models, and derive a mass matrix whose smaller eigenvalue has the interpretation of the inflaton mass-squared. This mass matrix may be used in the calculation of inflationary observables. In Sec.~\ref{sec:Results}, we compare the predictions for inflationary observable in a variety of models which differ in their kinetic terms, most of which already appear in the literature. We conclude in Sec.~\ref{sec:Conclusions}.

\section{Single-Field Effective Description}
\label{sec:1-field}

In this section we review the single-field description of spiral-inflation models with Lagrangian Eq.~(\ref{eq:L_SI}), and derive  a mass-matrix description relevant for computation of inflationary observables. We first review the role of the field-space metric in the single-field effective description of these models.

\subsection{From many fields to one}
Consider a model with real scalar fields $\phi^a$ in a background spacetime described by the metric $g_{\mu\nu}$.
During inflation we assume the spacetime is given by the flat Friedmann-Robertson-Walker (FRW) metric $g_{00}=1$, $g_{ij}=-a^2(t)\delta_{ij}$, where $i,j\in\{1,2,3\}$ and $t\equiv x^0$, but for now we allow an arbitrary time-dependent metric.  
The Lagrangian for the theory is,\begin{equation}
\sqrt{|g|}{\cal L}=\sqrt{|g|}\frac{1}{2}G_{ab}g^{\mu\nu}\partial_\mu\phi^a \partial_\nu\phi^b-\sqrt{|g|}V(\left\{\phi^a\right\}), \label{eq:L2field}
\end{equation}
where $G_{ab}(\{\phi^c\})$ in the kinetic terms defines the field-space metric, which is taken to be symmetric in $a\leftrightarrow b$.
Under a nonlinear field redefinition $\phi^a\rightarrow \tilde{\phi}^a\left(\left\{\phi^b\right\}\right)$,
 the Lagrangian transforms as,\begin{eqnarray}
\sqrt{|g|}{\cal L}&=&\sqrt{|g|}\frac{1}{2}G_{ab}\frac{\partial \phi^a}{\partial\tilde{\phi}^c}\frac{\partial \phi^b}{\partial\tilde{\phi}^d}g^{\mu\nu}\partial_\mu\tilde{\phi}^c \partial_\nu\tilde{\phi}^d-\sqrt{|g|}V\left(\phi^a(\{\tilde{\phi}^b\})\right) \\
&\equiv&\sqrt{|g|}\frac{1}{2}\tilde{G}_{cd}g^{\mu\nu}\partial_\mu\tilde{\phi}^c \partial_\nu\tilde{\phi}^d-\sqrt{|g|}V\left(\phi^a(\{\tilde{\phi}^b\})\right), 
\end{eqnarray}
which defines the transformed field-space metric as \begin{equation}
\tilde{G}_{cd}=G_{ab}\frac{\partial \phi^a}{\partial\tilde{\phi}^c}\frac{\partial \phi^b}{\partial\tilde{\phi}^d}.
\end{equation}
In this sense, the field-space metric transforms as a tensor under field transformations.
Locally one can redefine the fields so that the field-space metric is flat, $\tilde{G}_{cd}=\delta_{cd}$, but this can be done globally only if the field-space metric originally describes a flat field space.

In order to compare with a single-field description we consider the equations of motion. The equations of motion for the fields $\phi^a$ are,
\begin{equation}
\frac{1}{\sqrt{|g|}}\partial_\mu\left(G_{ab}(\{\phi\})\sqrt{|g|}g^{\mu\nu}\partial_\nu\phi^b\right)=-\frac{\partial V}{\partial \phi^a}+\frac{1}{2}g^{\mu\nu}
\frac{\partial G_{cb}}{\partial\phi^a}\partial_\mu\phi\,\partial_\nu\phi. \end{equation}
We will be interested in spatially uniform solutions to the equations of motion, so that the fields $\phi^a$ only have dependence on $t$. For these solutions, the equations of motion are 
\begin{equation}
\frac{1}{\sqrt{|g|}}\frac{d}{dt}\left(\sqrt{|g|}g^{00}G_{ab}\dot{\phi}^b \right)-\frac{1}{2}g^{00}\frac{\partial G_{cb}}{\partial\phi^a}\dot{\phi}^c\dot{\phi}^b=
-\frac{\partial V}{\partial \phi^a}, \label{eq:phiEOM}\end{equation}
where $\dot{\phi}^a\equiv d\phi^a/dt$.

Now suppose that the trajectory describing a solution to the equations of motion is known, parametrized by a parameter $I$ along the trajectory, so that along the given solution we have $\phi^a(I)$. For such a solution, the equations of motion determine the time dependence of $I$. 
Multiplying Eq.~(\ref{eq:phiEOM}) by $\phi^{a\,\prime}(I)$ gives, \begin{equation}
\frac{1}{\sqrt{|g|}}\phi^{a\,\prime}(I)\frac{d}{dt}\left(\sqrt{|g|}g^{00}G_{ab}\dot{\phi}^b \right)-\frac{1}{2}g^{00}G_{ab}'(I)\dot{\phi}^a\dot{\phi}^b=
-V'(I). \label{eq:phiEOM2}\end{equation}

Now choose $I$ to satisfy the field-space condition \begin{equation}
G_{ab}\phi^{a}\,'(I)\phi^{b}\,'(I)=1. \label{eq:affine}\end{equation}
This condition makes the parameter $I$ analogous to the invariant length, but in field space, and  will give $I$ the interpretation of a canonically normalized inflaton field, with kinetic term $\frac{1}{2}\dot{I}^2$.
A derivative of Eq.~(\ref{eq:affine}) with respect to $I$ gives, \begin{equation}
G_{ab}\,'(I)\phi^{a\,\prime}(I)\phi^{b\,\prime}(I)+2G_{ab}\phi^{a\,\prime\prime}(I)\phi^{b\,\prime}(I)=0.\end{equation}
Multiplying by $\dot{I}^2$, we have \begin{equation}
\frac{1}{2}G_{ab}\,'(I)\dot{\phi}^a\dot{\phi}^b=-G_{ab}\phi^{a\,\prime\prime}(I)\dot{\phi}^b\dot{I}. \label{eq:usefuleq1}
\end{equation}
Using Eq.~(\ref{eq:usefuleq1}), the equations of motion Eq.~(\ref{eq:phiEOM2}) become, \begin{equation}
\frac{1}{\sqrt{|g|}}\phi^{a\,\prime}(I)\frac{d}{dt}\left(\sqrt{|g|}g^{00}G_{ab}\dot{\phi}^b \right)+g^{00}G_{ab}\phi^{a\,\prime\prime}(I)\dot{\phi}^b\dot{I}=-V'(I). \label{eq:phiEOM3}\end{equation}
The first two terms in Eq.~(\ref{eq:phiEOM3}) combine to give a time derivative, \begin{equation}
\frac{1}{\sqrt{|g|}}\frac{d}{dt}\left(\sqrt{|g|}g^{00}G_{ab}\phi^{a\,\prime}(I)\phi^{b\,\prime}(I)\dot{I}\right)=-V'(I), \end{equation}
or using Eq.~(\ref{eq:affine}), \begin{equation}
\frac{1}{\sqrt{|g|}}\frac{d}{dt}\left(\sqrt{|g|}g^{00}\dot{I}\right)=-V'(I). \label{eq:2fieldto1field} \end{equation}
Together with the trajectory $\phi^a(I)$ that solves the equations of motion, a solution to Eq.~(\ref{eq:2fieldto1field}) then determines the time dependence of that trajectory.  Consequently, Eq.~(\ref{eq:2fieldto1field}) provides enough information to determine inflationary observables, as long as the fluctuations in the direction orthogonal to the trajectory are massive compared to $H^{-1}$ so that they are not produced during inflation.

The field-space parameter $I$ above plays the role of the inflaton  in the single-field description of any model with Lagrangian of the form Eq.~(\ref{eq:L2field}). The analysis above supposed that we knew the trajectory along a solution to the equations of motion.
Now suppose that we had instead imposed as a constraint that the fields lie on the trajectory $\phi^a(I)$. In Dante's inferno and spiral inflation models, the trajectory is approximately known due to the presence of a steep-walled trench in the potential. This is a holonomic constraint, as can be made explicit by inverting the expression for one of the fields, say $\phi^1(I)$ to give $I(\phi^1)$. We  assume that this inverse exists throughout the field trajectory. Then the remaining constraints are of the form $\phi^a-\phi^a\left(I(\phi^1)\right)=0$. Such constraints can be imposed either by Lagrange multipliers in the Lagrangian, or by simply replacing $\phi^a$ by $\phi^a(I)$ in the Lagrangian. We are left with a description of the theory in terms of the single field $I$. 

If we again choose $I$ to satisfy the  condition Eq.~(\ref{eq:affine}), then the Lagrangian Eq.~(\ref{eq:L2field}) constrained to a field-space trajectory takes the canonical form, \begin{equation}
\sqrt{|g|}{\cal L}_I=\sqrt{|g|}\left(\frac{1}{2}g^{00}\dot{I}^2-V(I)\right). \label{eq:LI}\end{equation}
The equations of motion that follow from this singe-field effective description are the same as Eq.~(\ref{eq:2fieldto1field}), which was derived in the multi-field description. This justifies the interpretation of the field $I$ as the canonical inflaton in these models. Note that the only assumption in the analysis of this section was that we knew the trajectory taken by the fields $\phi^a$, which in many inflation models is known by the presence of a steep-walled trench in the potential.

\subsection{Spiral Inflation Models and a Mass Matrix}
\label{sec:matrix}

At this stage we will focus on spiral inflation models, for which $G_{rr}=1$, $G_{\theta\theta}=r^2$, and $G_{r\theta}=G_{\theta r}=0$. The condition Eq.~(\ref{eq:affine}) defining the canonical inflaton field can be written \begin{equation}
dI^2=dr^2+r^2\,d\theta^2. \end{equation}

We suppose that the trajectory $r(\theta)$, approximately determined by the shape of the trench in the potential, is known.  At a given time, the inflaton direction in field space is specified by the unit vector
\begin{equation}
\hat{\mathbf{e}}_I=c_r\hat{\mathbf{e}}_r+c_\theta \hat{\mathbf{e}}_\theta,
\end{equation}
where \begin{equation}
c_r=\frac{dr}{dI}=\frac{r'(\theta)}{\sqrt{r^2+r'^2}},\ \ c_\theta=r\frac{d\theta}{dI}=\frac{r}{\sqrt{r^2+r'^2}}, \label{eq:crct}
\end{equation}
and the unit vectors $\hat{\mathbf{e}}_r$ and $\hat{\mathbf{e}}_\theta$ are the usual basis vectors in polar coordinates, which in a Cartesian coordinate system with $x=r \cos\theta,\ y=r\sin\theta$ have components $\hat{\mathbf{e}}_r=\cos\theta\,\hat{\mathbf{e}}_x+\sin\theta\,\hat{\mathbf{e}}_y$,
$\hat{\mathbf{e}}_\theta=-\sin\theta\,\hat{\mathbf{e}}_x+\cos\theta\,\hat{\mathbf{e}}_y$.
In spiral inflation models the field evolution is mostly in the $\hat{\mathbf{e}}_\theta$ direction. In order to compare with a mass matrix description, as in Ref.~\cite{Barenboim:2014vea}, we make the approximation that the trajectory is nearly circular, and set to zero $c_r\,'(\theta),\ c_\theta\,'(\theta)$, which is a good approximation for typical parameter choices in these models as we will confirm numerically in Sec.~\ref{sec:Results}.

The slow-roll parameters, and consequently inflationary observables, depend on derivatives of the potential with respect to the canonically normalized inflaton field. In multi-field models this is a directional derivative (which for comparison with the previous section is simply the chain rule with Eq.~(\ref{eq:crct})):
\begin{equation}
\frac{dV}{dI}=(\hat{\mathbf{e}}_I\cdot \nabla)V=c_r\,\partial_r V+c_\theta/r\,\partial_\theta V,\end{equation}
where $\nabla V$ is the gradient in polar coordinates, $\nabla V=\partial_r V\,\hat{\mathbf{e}}_r+1/r\,\partial_\theta V\,\hat{\mathbf{e}}_\theta$.
The derivative $dV/dI$  determines the slow-roll parameter $\epsilon$ defined by \begin{equation}
\epsilon=\frac{M_*^2}{2}\left(\frac{V'(I)}{V}\right)^2,\end{equation}
where $M_*=2.4\times 10^{18}$ GeV is the reduced Planck mass.  Noting that \begin{equation}
\frac{d\hat{\mathbf{e}}_r}{d\theta}=\hat{\mathbf{e}}_\theta,\ \ \frac{d\hat{\mathbf{e}}_\theta}{d\theta}=-\hat{\mathbf{e}}_r, \end{equation}
we have
\begin{eqnarray}
\frac{d^2V}{dI^2}&=&\frac{d}{dI}\left(\hat{\mathbf{e}}_I\cdot \nabla\right)V \\
&=&\frac{d\hat{\mathbf{e}}_I}{dI}\cdot\nabla V+\hat{\mathbf{e}}_I\cdot \frac{d}{dI}(\nabla V) \\
&=&\frac{d\theta}{dI}(c_r\hat{\mathbf{e}}_\theta-c_\theta\hat{\mathbf{e}}_r)\cdot(\partial_rV \hat{\mathbf{e}}_r+\frac{1}{r}\partial_\theta V \hat{\mathbf{e}}_\theta)
 \nonumber \\ &&+\hat{\mathbf{e}}_I\cdot\left[\left((\hat{\mathbf{e}}_I\cdot\nabla)\partial_r V\right)\hat{\mathbf{e}}_r+\left((\hat{\mathbf{e}}_I\cdot\nabla)\frac{1}{r}\partial_\theta V\right)\hat{\mathbf{e}}_\theta\right] 
\\ &&+\hat{\mathbf{e}}_I\cdot\left[\partial_r V\frac{d\hat{\mathbf{e}}_r}{dI}+\frac{1}{r}\partial_\theta V\frac{d\hat{\mathbf{e}}_\theta}{dI}\right]
 \nonumber  \label{eq:Vpp}\end{eqnarray}
 
   Eq.~(\ref{eq:Vpp}) can be simplified using \begin{equation}
  \frac{d\theta}{dI}=\frac{c_\theta}{r}, \end{equation}
  yielding \begin{eqnarray}
  \frac{d^2 V}{dI^2}&=&c_r^2\partial_r^2V+2\frac{c_r c_\theta}{r}\partial_r\partial_\theta V+\frac{c_\theta^2}{r^2}\partial_\theta^2 V-\frac{c_rc_\theta}{r^2}\partial_\theta V \\
  &=&\left(\begin{array}{cc}c_r,&c_\theta\end{array}\right)\left(\begin{array}{cc}
    \partial_r^2V&\frac{1}{r}\partial_r\partial_\theta V -\frac{1}{2r^2}\partial_\theta V \\
  \frac{1}{r}  \partial_r\partial_\theta V -\frac{1}{2r^2}\partial_\theta V &\frac{1}{r^2}\partial_\theta^2 V\end{array}\right)\left(\begin{array}{c}c_r\\c_\theta\end{array}\right).
\end{eqnarray}
We can now identify the mass matrix appropriate for calculation of inflationary observables, 
\begin{equation}
M^2_{r\theta}=\left(\begin{array}{cc}
    \partial_r^2V&\frac{1}{r}\partial_r\partial_\theta V -\frac{1}{2r^2}\partial_\theta V \\
   \frac{1}{r} \partial_r\partial_\theta V -\frac{1}{2r^2}\partial_\theta V &\frac{1}{r^2}\partial_\theta^2 V\end{array}\right).
\label{eq:Msq_rt}    \end{equation}
In particular, the slow-roll parameter $\eta$ is defined as, \begin{equation}
\eta=M_*^2\frac{V''(I)}{V},\end{equation}
which may be calculated directly in the single-field effective description, or else (to good approximation) as the smaller eigenvalue of the mass matrix $M^2_{r\theta}$.

We note  that the mass matrix $M^2_{r\theta}$ differs from the mass matrix of Refs.~\cite{Barenboim:2014vea,Barenboim:2015zka,Barenboim:2015lla} in the off-diagonal terms, which explains differences in the results of this paper and those of some earlier papers.\footnote{We are grateful to Gabriela Barenboim and Wan-Il Park for discussion on this point.} In particular, by identifying successive derivatives in the $\hat{\mathbf{e}}_r$ and $\hat{\mathbf{e}}_\theta$ directions as $\partial_r$ and $\partial_\theta/r$, respectively, the mass matrix of Refs.~\cite{Barenboim:2014vea,Barenboim:2015zka,Barenboim:2015lla} neglects the $1/(2r^2)\partial_\theta V$ term in the off-diagonal elements of Eq.~(\ref{eq:Msq_rt}). 
It is perhaps worthwhile therefore to discuss  other mass matrices whose eigenvalues  are not directly related to derivatives with respect to the inflaton in the single-field description. To that effect we will introduce some well motivated straw-man mass matrices in spiral inflation models, and describe their physical interpretation in relation to the inflaton dynamics.

Rather than begin with the  field-space variables $r$ and $\theta$ in spiral inflation models, which have noncanonical kinetic terms, one might have instead considered beginning with field-space variables $x^1\equiv r\cos\theta$, $x^2\equiv r\sin\theta$, in which case the kinetic terms are canonical and one can define the  mass matrix  $(M^2_{{\rm Cartesian}})_{ij}\equiv \partial_i\partial_j V[r(x,y),\theta(x,y)]$, where $\partial_i\equiv\partial/\partial x^i$. This mass matrix, evaluated at a point in field space, determines the quadratic terms in a Taylor expansion of the potential about that point.  Then transforming to the polar variables in the neighborhood of that point, $(dx,dy)^T\rightarrow(dr,r\,d\theta)^T=R(\theta)(dx,dy)^T$, where $R(\theta)$ is the 2$\times$2 rotation matrix with angle $\theta$, gives the mass matrix $\tilde{M}^2_{{\rm Cartesian}}$, where \begin{equation}
\tilde{M}^2_{{\rm Cartesian}}=R(\theta)M^2R^{-1}(\theta)=\left(\begin{array}{cc}
    \partial_r^2V&\frac{1}{r}\partial_r\partial_\theta V -\frac{1}{r^2}\partial_\theta V \\
   \frac{1}{r} \partial_r\partial_\theta V -\frac{1}{r^2}\partial_\theta V &\frac{1}{r^2}\partial_\theta^2 V+\frac{1}{r}\partial_r V\end{array}\right),
    \label{eq:Msq_Cartesian}\end{equation}
so that a Taylor expansion of the potential {\em in Cartesian coordinates} about a point $(r_0,\theta_0)$  has quadratic part,
    \begin{equation}
V(r,\theta)=\cdots+\left(\begin{array}{cc}dr,& r\,d\theta\end{array}\right)\left(\begin{array}{cc}
    \partial_r^2V&\frac{1}{r}\partial_r\partial_\theta V -\frac{1}{r^2}\partial_\theta V \\
   \frac{1}{r} \partial_r\partial_\theta V -\frac{1}{r^2}\partial_\theta V &\frac{1}{r^2}\partial_\theta^2 V+\frac{1}{r}\partial_r V\end{array}\right)\left(\begin{array}{c}dr\\r\,d\theta\end{array}\right)+\cdots,\end{equation}
where $dr=(r-r_0)$, $d\theta=(\theta-\theta_0)$.
The matrix $\tilde{M}^2_{{\rm Cartesian}}$  is also closely related to the matrix of covariant derivatives in polar coordinates, \begin{equation}
M^2_{{\rm cov}\,ab}=D_aD_bV=\partial_a\partial_bV-\Gamma^c_{ab}\partial_cV,\end{equation}
except that $\theta$ components have been rescaled by $1/r$ in $\tilde{M}^2_{{\rm Cartesian}}$ to transform to the basis $(dr, r\,d\theta)$ from $(dr,d\theta)$.  Here, $\Gamma^c_{ab}$ is the Christoffel symbol in field space, with nonvanishing components,
\begin{eqnarray}
\Gamma^r_{\theta\theta}&=& -r, \\
\Gamma^\theta_{r\theta}=\Gamma^\theta_{\theta r}&=&1/r.
    \end{eqnarray}

The eigenvectors of the various mass matrices   described above are numerically similar along the trench defined by $\partial_r V=0$ in the models considered in this paper. The eigenvalues of the mass matrices, however are quite different. This is illustrated in Fig.~\ref{fig:matrix} in a numerical example of Sec.~\ref{sec:Results}. 

To summarize this section, with knowledge of the trajectory describing the evolution of  fields constrained to follow a steep-walled trench during inflation, one can define a single-field effective description in terms of a potential $V(I)$ in terms of a canonically normalized inflaton field $I$. The single-field description allows for straightforward computation of inflationary observables, and is the usual procedure for calculation of observables in multi-field models. A mass matrix relating the single-field and multi-field descriptions may be constructed, and differs significantly from the mass matrix as usually defined if the direction of field evolution varies significantly during inflation, as in spiral inflation models.

\section{Results}
\label{sec:Results}
We consider theories with both canonical and non-canonical kinetic terms in this section. We use units of the reduced Planck mass $M_*= 2.4\times 10^{18}$~GeV throughout. Respectively, the Lagrangians are of the  form Eq.~(\ref{eq:L_DI}) and Eq.~(\ref{eq:L_SI}),
where $V(r,\theta)=W(r)+\Lambda^4\left[1-\cos\left((\frac{r}{f})^n-\theta\right)\right]$. 
The inflaton field is defined so that along a trajectory $(r(t),\ \theta(t))$ the field is canonically normalized. Recall that in the Dante's inferno-type model the fields $r$ and $\theta$ are canonically normalized, and in spiral inflation models the fields are non-canonically normalized. In these cases, respectively, the inflaton field $I(t)$ satisfies
\begin{eqnarray}
dI_{\rm C}=\frac{\dot{r}}{\sqrt{\dot{r}^2+\dot{\theta}^2}}dr+\frac{\dot{\theta}}{\sqrt{\dot{r}^2+\dot{\theta}^2}}d\theta \,\, , \nonumber \\
dI_{\rm NC}=\frac{\dot{r}}{\sqrt{\dot{r}^2+r^2\dot{\theta}^2}}dr+\frac{r\dot{\theta}}{\sqrt{\dot{r}^2+r^2\dot{\theta}^2}}rd\theta \,\, .
\label{eq:inf}
\end{eqnarray}
In both cases, the trajectory closely follows the bottom of the trench defined by $\partial V(r,\theta)/\partial r=0$, or
\beq
\sin\left((\frac{r}{f})^n-\theta\right)=-\frac{f^n}{n \Lambda^4}W'(r) r^{1-n} \,\, .
\label{eq:trench}
\enq
We denote the trajectory by $r(\theta)$. Eq.~(\ref{eq:inf}) can be restated as
\begin{eqnarray}
dI_{\rm C}=\frac{r'}{\sqrt{r'^2+1}}dr+\frac{1}{\sqrt{r'^2+1}}d\theta=\sqrt{r'^2+1} \, d\theta \,\, , \nonumber \\
dI_{\rm NC}=\frac{r'}{\sqrt{r'^2+r^2}}dr+\frac{r}{\sqrt{r'^2+r^2}}rd\theta=\sqrt{r'^2+r^2} \, d\theta \,\, .
\label{eq:inf2}
\end{eqnarray}
The derivative of $V$ with respect to $I$ becomes
\begin{eqnarray}
\frac{dV}{dI_{\rm C}}=\frac{1}{\sqrt{r'(\theta)^2+1}}\frac{d V(r(\theta),\theta)}{d \theta} \,\, , \nonumber \\
\frac{dV}{dI_{\rm NC}}=\frac{1}{\sqrt{r'(\theta)^2+r(\theta)^2}}\frac{d V(r(\theta),\theta)}{d \theta} \,\, .
\label{eq:infdd}
\end{eqnarray}
We normally work in the region where $r'(\theta)\ll 1$ in the canonical case, and and $r'(\theta)\ll r$ in the non-canonical case. Then, Eq.~(\ref{eq:infdd}) can be approximated by
\begin{eqnarray}
\frac{dV}{dI_{\rm C}}\approx \frac{d V(r(\theta),\theta)}{d \theta} \,\, , \nonumber \\
\frac{dV}{dI_{\rm NC}}\approx \frac{1}{r(\theta)}\frac{d V(r(\theta),\theta)}{d \theta}\,\, .
\label{eq:infdda}
\end{eqnarray}
The slow-roll parameters can now be calculated by
\beq
\epsilon \equiv \frac{M_{*}^2}{2}\left(\frac{V'(I)}{V}\right)^2 \,\, , \quad
\eta \equiv M_{*}^2 \frac{V''(I)}{V} \,\, , \quad
\gamma \equiv M_{*}^4 \frac{V'(I)\,V'''(I)}{V^2} \,\, .
\label{eq:srp}
\enq
The inflationary observables are then given by
\beq
\tilde{r}=\left[16\epsilon\right]_{I=I_i} \, , \quad
n_s=\left[1+2\eta-6\epsilon\right]_{I=I_i} \, , \quad
\Delta_R^2=\left[\frac{V}{24 \pi^2 \epsilon}\right]_{I=I_i} \, , \quad
n_r=\left[16\epsilon \eta-24\epsilon^2-2\gamma\right]_{I=I_i} \, ,
\enq
where $I_i$ is the value of the inflaton field at the time when the observed inflationary perturbations were created, which in most models is 50-60 $e$-folds before the end of inflation, but is sensitive to the details of reheating after inflation.
The observable $\tilde{r}$ is the ratio of the tensor to scalar amplitude, where we use the unconventional tilde over $r$ to distinguish the observable from  the field $r$ in these models. The other observables are the scalar tilt $n_s$; the scalar amplitude $\Delta_R^2$,  also denoted $A_s$; and the running of the scalar tilt $n_r$. Definitions in terms of the CMB spectrum are available in many places, for example in the Planck 2015 results
papers \cite{Ade:2015lrj}.

The number of $e$-folds is given by
\beq
N_e=\int_{I_i}^{I_f} \frac{V}{V'(I)} dI \,\, .
\label{eq:ne}
\enq
In our numerical analysis we determine the initial point of inflation  by fixing $n_s=0.96$ and $\Delta_R^2=2.2\times 10^{-9}$, close to the values measured by the Planck experiment \cite{Ade:2015lrj}, $n_s=0.9655\pm0.0062$, $\ln(10^{10}\Delta_R^2)=3.089\pm0.036$. The current experimental
constraint on $n_r$ is based on the Planck measurement, $n_r=-0.003\pm0.015$ \cite{Ade:2015lrj}. The end of inflation occurs when either
\beq
\left[\epsilon\right]_{I=I_f}=1 ,
\label{eq:endI}
\enq
or when the potential reaches a hybrid-inflation-type instability as in the Dante's waterfall model.
Two types of $W(r)$ are studied in the following sections and their corresponding single-field approximations are compared with the full theory.

\subsection{$\lambda r^p$}
We first consider $W(r)=\lambda r^p$. The trench equation Eq.~(\ref{eq:trench}) becomes
\beq
\sin\left((\frac{r}{f})^n-\theta\right)=-\frac{p \lambda f^n}{n \Lambda^4}r^{p-n} \,\, .
\label{eq:trenchl}
\enq
We consider the case that during inflation the magnitude of the right-hand side of Eq.~(\ref{eq:trenchl}) is $\ll 1$, corresponding to a steep-walled trench, so that Eq.~(\ref{eq:trenchl}) can be solved by
\beq
\theta=\frac{r^n}{f^n}+\frac{p \lambda f^n}{n \Lambda^4}r^{p-n}
\label{eq:trenchlr}
\enq
up to a constant phase. If we choose parameters so that the second term on the right-hand side is negligible, Eq.~(\ref{eq:trenchlr}) reduces to $r=f\theta^{\frac{1}{n}}$, and away from the global minimum of the potential we have $V(r(\theta),\theta) \approx W(r(\theta))=\lambda f^p\theta^{\frac{p}{n}}$. From Eq.~(\ref{eq:inf2}), we have 
\begin{eqnarray}
dI_{\rm C}\approx d\theta \,\, , \nonumber \\
dI_{\rm NC}\approx f \theta^{\frac{1}{n}} \, d\theta \,\, .
\label{eq:dI}
\end{eqnarray}
The single-field description of the potential in this approximation is therefore given by the potential,
\begin{eqnarray}
V_{\rm C}(I) \sim I^{\frac{p}{n}} \,\, , \nonumber \\
V_{\rm NC}(I) \sim I^{\frac{p}{n+1}} \,\, .
\label{eq:VI}
\end{eqnarray}
We work through the $(p=4,\, n=1,\,2)$ case for illustration.

\subsubsection{$p=4, \,\, n=1$} 
First we show the predictions of the observables from the single-field approximation. Using Eqs.~(\ref{eq:infdda})$\text{--}$(\ref{eq:endI}), we analyze theories with both canonical and non-canonical kinetic terms, as earlier. For $(p,n)=(4,1)$, Eq.~(\ref{eq:trenchlr}) is now $\theta=\frac{r}{f}+\frac{4 \lambda f}{\Lambda^4}r^{3}$. Assuming the second term on the right-hand-side is negligible, we get that the trench follows $r(\theta) \approx f \theta$ thus $V(r(\theta), \theta) \approx W(r(\theta))=\lambda f^4 \theta^4$. We determine the initial and final point of inflation in field space  by fixing $n_s=0.96$ and $\left[\epsilon\right]_{\theta=\theta_f}=1$. 
Note that $n_s$ and $\epsilon$ are not sensitive to the overall scale in the potential while $\Delta_R^2$ is, so $\Delta_R^2$ can  be controlled by rescaling the potential. Fixing observables this way, the model then predicts the number of $e$-folds during inflation and the ratio of tensor to scalar amplitudes $\tilde{r}$. The results are given in Table~I.
\begin{table}[htbp]
\label{tab:l1r}
\begin{center}
\begin{tabular} {|ccc rccccccccccc|}
\hline\hline
&&& && $\theta_i$ && $\theta_f$ && $\tilde{r}$ && $N_e$&&$V(I)$&\\[0.5ex]
\hline
&$C$& &&& $10\sqrt{6}$ && $2\sqrt{2}$ && $0.2133$ && $74$&&$\sim I^4$&\\
\hline
&$NC$& &&& $(\frac{800}{f^2})^{\frac{1}{4}}$ && $(\frac{8}{f^2})^{\frac{1}{4}}$ && $0.16$ && $49.5$&&$\sim I^2$&\\
\hline
\end{tabular}
\caption{Observables from the single-field approximation for the $(p,n)=(4,1)$ model, fixing $n_s=0.96$ and $\left[\epsilon\right]_{\theta=\theta_f}=1$.}
\end{center}
\end{table}

\begin{figure}[ht]
\subfigure{\includegraphics[width = 0.495\textwidth]{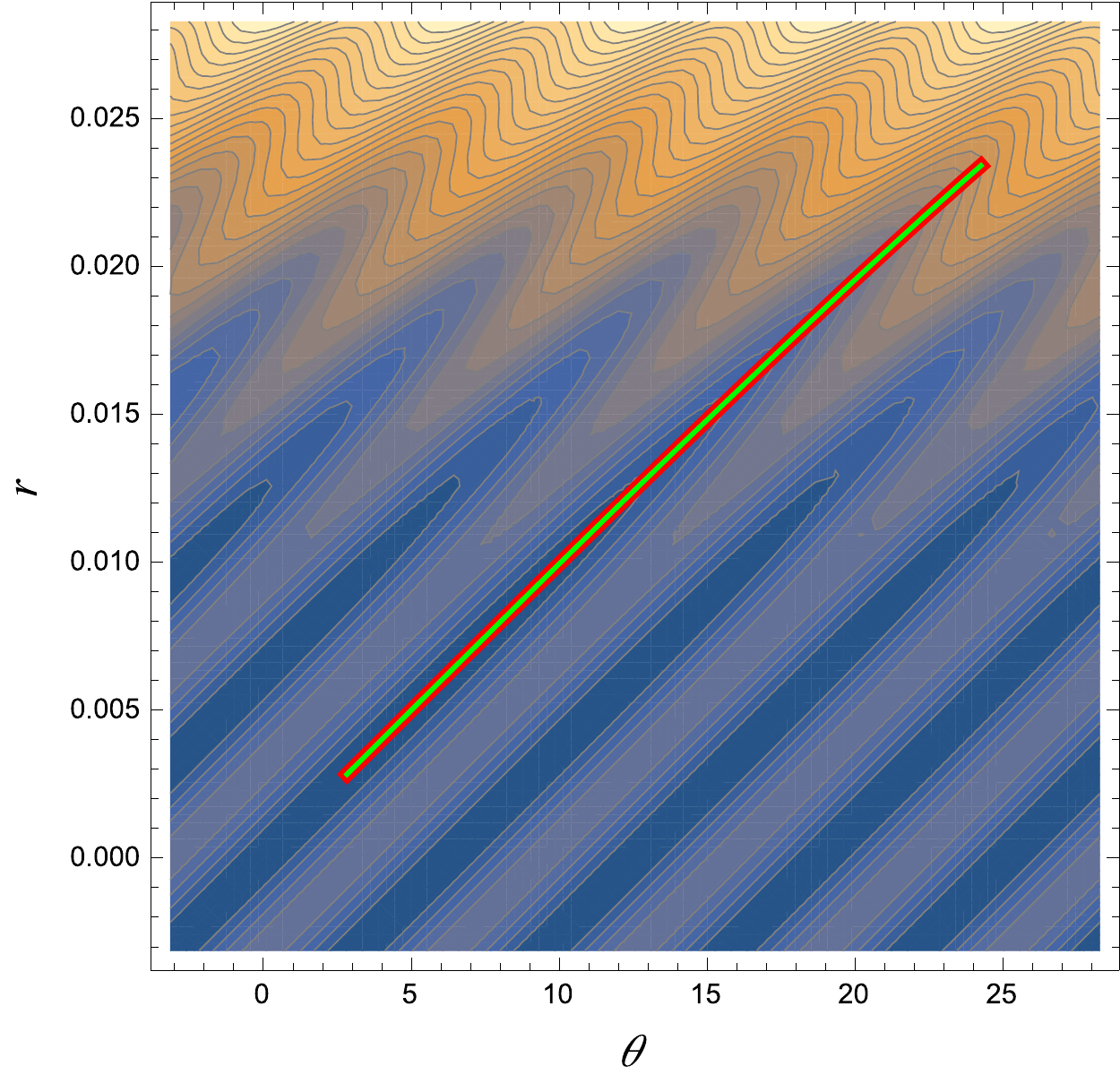}}
\subfigure{\includegraphics[width = 0.495\textwidth]{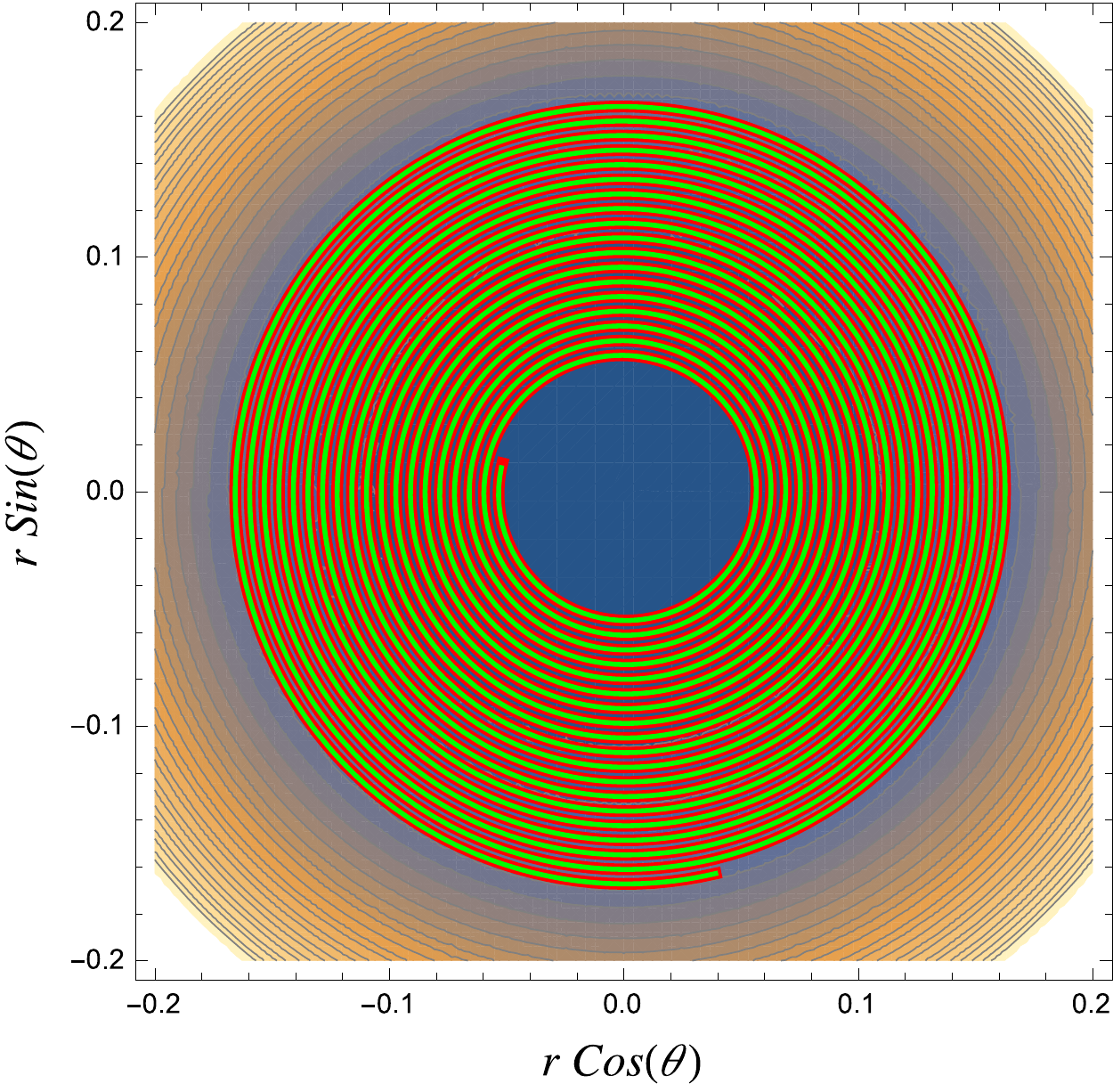}}
\caption{Contour plot of the potential for $p=4, \,\, n=1$. The canonical case is plotted on the left, the non-canonical case is plotted on the right. The red line indicates the bottom of the trench. The inflationary trajectory is shown by the green line.}
\label{fig:r41}
\end{figure}

Working numerically in the complete two-field model, we find the following examples of parameter sets, in units  $M_*$=$1$.  With $(\lambda, \, \Lambda^4, \, f)_{\rm C}=(0.02025, \, 1.377\times 10^{-9}, \, 0.001)$ and $(\lambda, \, \Lambda^4, \, f)_{\rm NC}=(6.525\times 10^{-6}, \, 1.68\times 10^{-10}, \, 0.001)$, we get that $(\tilde{r}, \, n_s, \,  n_r, \, \Delta_R^2, \, N_e)_{\rm C}=(0.2012, \, 0.96, \, -4.88\times 10^{-4}, \, 2.2\times 10^{-9}, \, 73.87)$ and $(\tilde{r}, \, n_s, \, n_r, \, \Delta_R^2, \, N_e)_{\rm NC}=(0.1593, \, 0.96, \, -7.97\times 10^{-4}, \, 2.2\times 10^{-9}, \, 49.52)$. Note that in the non-canonical case the coupling $\lambda$ is driven to be nonperturbative and the perturbative analysis is not valid, but for the purpose of comparison with the single-field description we treat this case classically.The results match well with those from Table I, derived from the single-field approximation. The dynamical solutions to the equations of motion Eq.~(\ref{eq:phiEOM}) are plotted in Fig~\ref{fig:r41}. 

Note that the nontrivial field-space metric in the non-canonical case has the consequence of reducing both the number of $e$-folds and $\tilde{r}$. However, this model is  ruled out by the large values of $\tilde{r}>0.11$ \cite{Ade:2015lrj} and $N_e>60$ in the canonical case, and the large value of $\tilde{r}$ in the non-canonical case.  

\begin{figure}[ht]
\subfigure{\includegraphics[width = 0.495\textwidth]{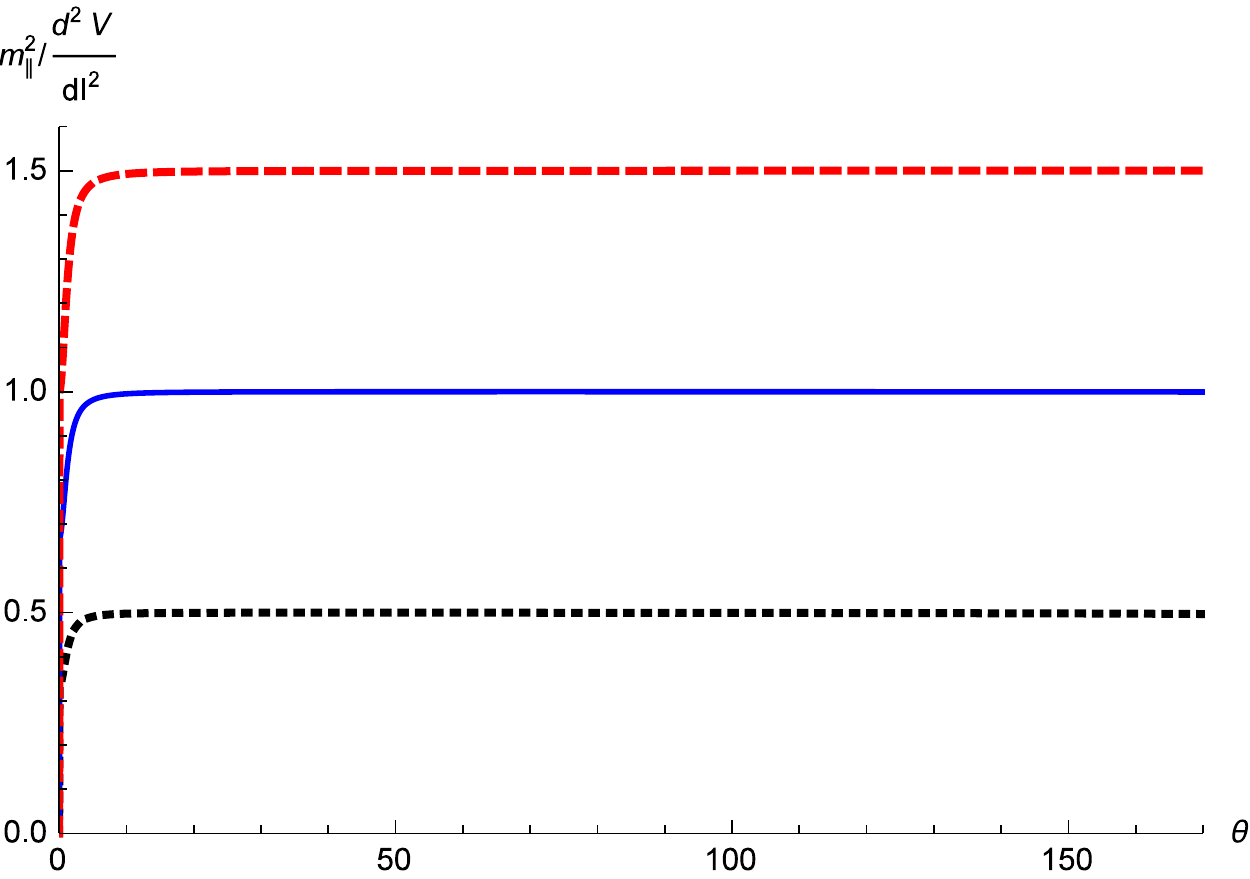}}
\subfigure{\includegraphics[width = 0.495\textwidth]{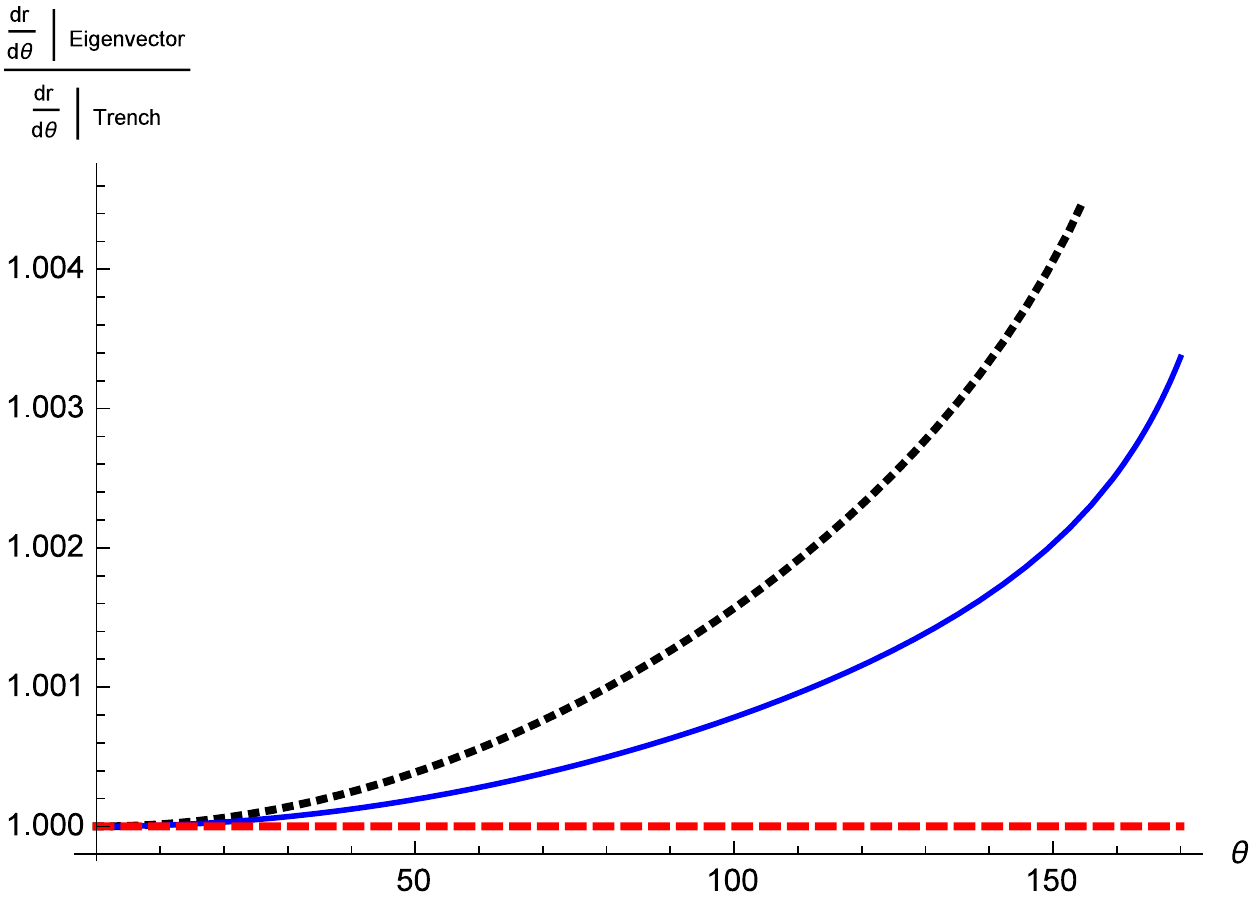}}
\caption{The solid blue line,  dotted black line, and  dashed red line correspond to our mass matrix Eq.~(\ref{eq:Msq_rt}), the Cartesian mass matrix Eq.~(\ref{eq:Msq_Cartesian}), and the mass matrix of Refs.~\cite{Barenboim:2014vea,Barenboim:2015zka,Barenboim:2015lla}, respectively. The lower eigenvalue of each matrix, indicated as $m^2_\parallel$ in units of $d^2V/dI^2$, is plotted along the trench in the left graph. The corresponding eigenvector's slope is shown on the right, compared to that of the trench.}
\label{fig:matrix}
\end{figure}

For the non-canonical case, the eigenvalues and eigenvectors of the three different matrices discussed in Sec.~\ref{sec:matrix} are shown in Fig~\ref{fig:matrix}. Note that the eigenvectors are similar for all three mass matrices, but the eigenvalues disagree. The solid blue line corresponds to the mass matrix of Eq.~(\ref{eq:Msq_rt}), and the smaller eigenvalue of this matrix agrees with the second derivative of the potential along the inflaton direction. Hence, diagonalizing this mass matrix allows for calculation of observables that depend on that second derivative, although it is simpler to 
work with the single-field effective description.

\subsubsection{$p=4, \,\, n=2$}
 For $(p,n)=(4,2)$, Eq.~(\ref{eq:trenchlr}) gives $r=\alpha \, \sqrt{\theta}$, where $\alpha=(\frac{1}{f^2}+\frac{4 \lambda f^2 }{2 \Lambda^4})^{-\frac{1}{2}}$. Thus $V(r(\theta), \theta) \approx W(r(\theta))=\lambda \alpha^4 \theta^2$. Following the analysis of the previous section, the results are given in Table II.
\begin{table}[htbp]
\begin{center}
\begin{tabular} {|ccc rccccccccccc|}
\hline\hline
&&& && $\theta_i$ && $\theta_f$ && $\tilde{r}$ && $N_e$&&$V(I)$&\\[0.5ex]
\hline
&$C$& &&& $10\sqrt{2}$ && $\sqrt{2}$ && $0.16$ && $49.5$&&$\sim I^2$&\\
\hline
&$NC$& &&& $(\frac{250}{\alpha^2})^{\frac{1}{3}}$ && $(\frac{2}{\alpha^2})^{\frac{1}{3}}$ && $0.128$ && $41.33$&&$\sim I^{\frac{4}{3}}$&\\
\hline
\end{tabular}
\caption{Observables from the single-field approximation for the $(p,n)=(4,2)$ model, fixing $n_s=0.96$ and $\left[\epsilon\right]_{\theta=\theta_f}=1$. }
\end{center}
\label{tab:l2r}
\end{table}

Numerical results of the complete two-field models follow. With $(\lambda, \, \Lambda^4, \, f)_{\rm C}=(27.5, \, 8.8\times 10^{-10}, \, 0.001)$ and $(\lambda, \, \Lambda^4, \, f)_{\rm NC}=(0.0105, \, 2.1\times 10^{-11}, \, 0.001)$, we get that $(\tilde{r}, \, n_s, \,  n_r, \, \Delta_R^2, \, N_e)_{\rm C}=(0.1578, \, 0.96, \, -7.83\times 10^{-4}, \, 2.2\times 10^{-9}, \, 49.71)$ and $(\tilde{r}, \, n_s, \, n_r, \, \Delta_R^2, \, N_e)_{\rm NC}=(0.128, \, 0.96, \, -9.6\times 10^{-4}, \, 2.2\times 10^{-9}, \, 41.33)$. Note that in the canonical case the coupling $\lambda$ is driven to be nonperturbative and the perturbative analysis is not valid, but for the purpose of comparison with the single-field description we treat this case classically. The results match relatively well with those from single-field approximations. Note that, again, the non-canonical kinetic term leads to a reduced $\tilde{r}$ and $N_e$.

\begin{figure}[ht]
\subfigure{\includegraphics[width = 0.495\textwidth]{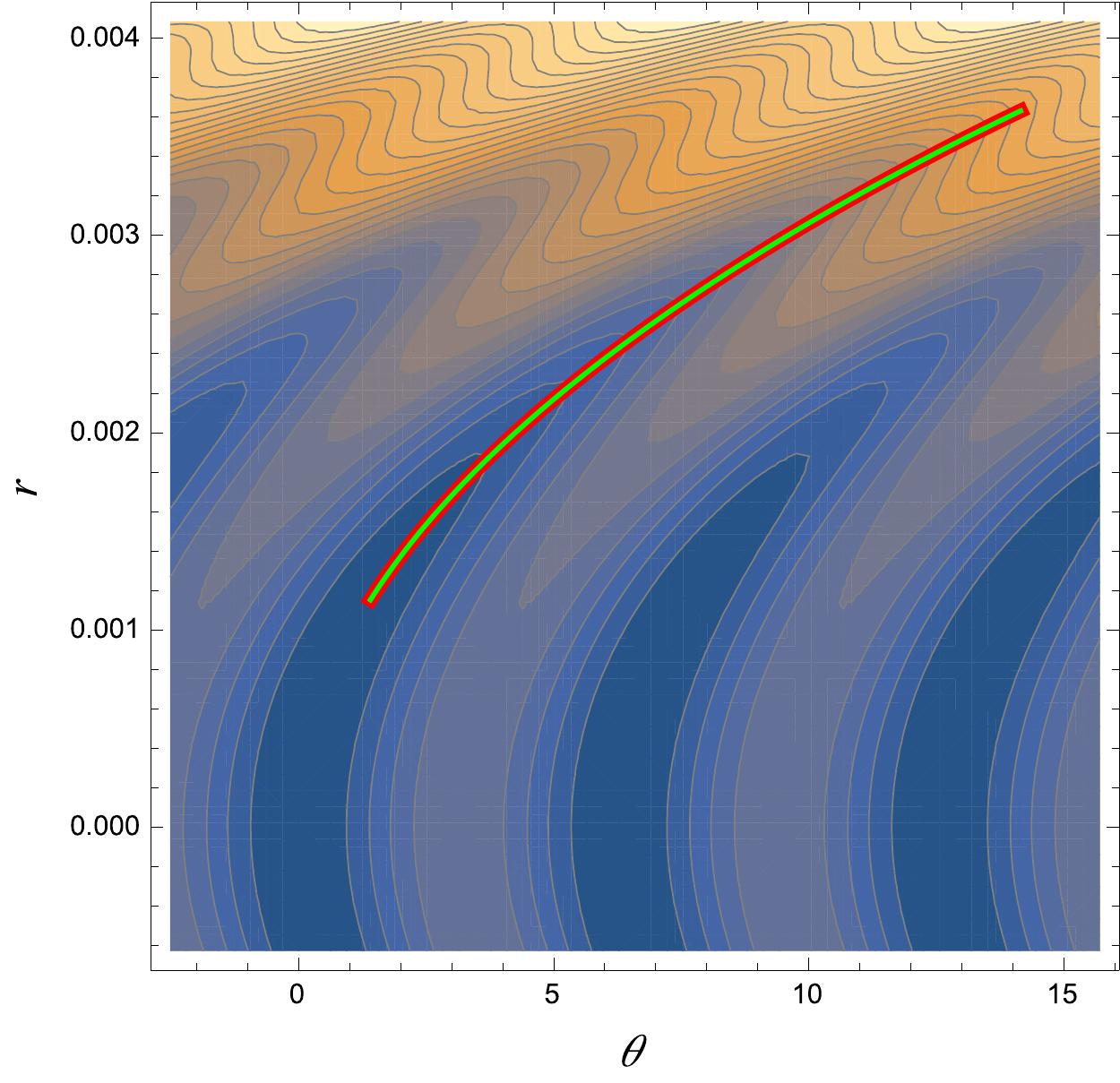}}
\subfigure{\includegraphics[width = 0.495\textwidth]{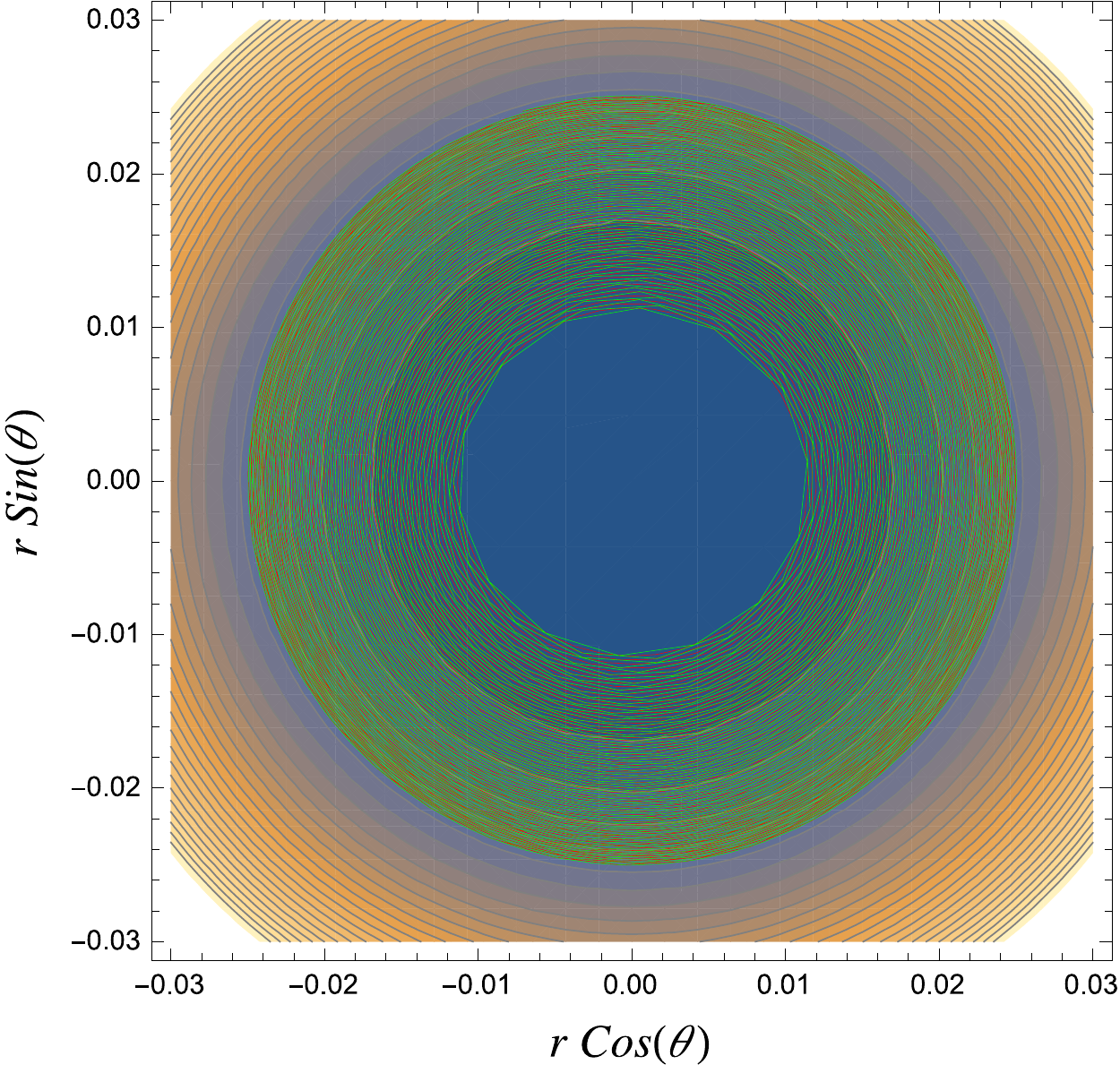}}
\caption{Contour plot of the potential for $p=4, \,\, n=2$. The canonical case is plotted on the left, the non-canonical case is plotted on the right. The red line indicates the bottom of the trench. The inflationary trajectory is shown by the green line.}
\label{fig:r42}
\end{figure}

We also notice that the $(4, \, 2)_{\rm C}$ model gives similar numerical predictions to the $(4,1)_{\rm NC}$ model. More generally, from Eq.~(\ref{eq:VI}) we see that the $(p, \, n+1)_{\rm C}$ model and $(p, \, n)_{\rm NC}$ model have the same single-field approximation. This is a type of duality between inflation models. The dynamical solutions are plotted in Fig~\ref{fig:r42}.

\subsubsection{$\lambda I^p$ with the desired observables}
We see from the above that the viability of these models is sensitive only to the power $p$ in the single-field effective description, as long as the single-field description is valid. Here we assume a simple potential $V(I)=\lambda I^p$ in the single-field description and work out the value of $p$ that would reproduce desired observables with the inflation process spanning $60$ $e$-folds. Using Eqs.~(\ref{eq:srp})$\text{--}$(\ref{eq:endI}) and fixing $n_s=0.96$, one gets $N_e=\frac{49}{4}p+25$. Imposing $N_e=60$, we have $p=20/7$. $\tilde{r}$ is then calculated to be $0.188$. Other observables may be calculated or fixed as in the earlier analysis. 

\subsection{$-\frac{1}{2}m^2r^2+\frac{\lambda}{4} r^4+\frac{m^4}{4\lambda}$}
Now we consider the potential $W(r)=-\frac{1}{2}m^2r^2+\frac{\lambda}{4} r^4+\frac{m^4}{4\lambda}$ as in Dante's waterfall. Eq.~(\ref{eq:trench}) becomes
\beq
\sin\left((\frac{r}{f})^n-\theta\right)=\frac{f^n}{n \Lambda^4}(m^2r^{2-n}-\lambda r^{4-n}) \,\, .
\label{eq:trenchW}
\enq
We again consider the $n= 1, \, 2$ cases for this potential and assume that the inflationary system starts near the origin where $-\frac{1}{2}m^2 r^2$ dominates over $\frac{\lambda}{4} r^4$.

\subsubsection{$n=1$}
With $n=1$, Eq.~(\ref{eq:trenchW}) reduces to $r=\alpha \theta$ if we neglect the $\lambda$ term, where $\alpha=(\frac{1}{f}-\frac{f m^2}{\Lambda^4})^{-1}$. For a stable trench to exist, the trench equation Eq.~(\ref{eq:trenchW}) should be solvable; however, over a range of $r$ a solution might not exist, depending on the model parameters \cite{Carone:2014cta,Barenboim:2014vea}. The canonical case is analyzed in Ref.~\cite{Carone:2014cta}, where a viable parameters space is found with inflation ending as in hybrid inflation. For the non-canonical case,
with $(m, \, \lambda, \, \Lambda^4, \, f)=(8.88\times 10^{-4}, \, 33.5 \, m^2, \, 7.2\times 10^{-5}\, m^2, \, 0.001)$, we have $(\tilde{r}, \, n_s, \, n_r, \, \Delta_R^2, \, N_e)=(0.1608, \, 0.96, \, -7.55\times 10^{-4}, \, 2.2\times 10^{-9}, \, 49.77)$. The end of inflation happens when $\left[\epsilon\right]_{\theta=\theta_f}=1$ in this example. Alternatively, if we change the $\Lambda^4$ to be $6.65\times 10^{-5}\, m^2$ in the above example, this model becomes a hybrid model as Dante's waterfall and inflation ends when the trench loses stability. The observables become $(\tilde{r}, \, n_s, \, n_r, \, \Delta_R^2, \, N_e)=(0.1610, \, 0.96, \, -7.52\times 10^{-4}, \, 2.2\times 10^{-9}, \, 21.22)$. 

We note that in the Dante's waterfall model the ratio of tensor to scalar amplitudes $\tilde{r}$ was found to be typically small with $\tilde{r}<0.03$. The noncanonical kinetic term in the  spiral inflation models above would predict larger values of $\tilde{r}$ but smaller $N_e$ than in the Dante's waterfall model, and it is challenging to find a viable parameter space in this class of spiral inflation models.

\subsubsection{$n=2$}
With $n=2$, Eq.~(\ref{eq:trenchW}) leads to $r=\alpha \sqrt{\theta+\beta}$, where $\alpha=(\frac{1}{f^2}+\frac{\lambda f^2}{2 \Lambda^4})^{-\frac{1}{2}}$ and $\beta=\frac{f^2m^2}{2\Lambda^4}$. We define a new field $\theta' \equiv \theta+\beta$, thus $V(r(\theta'))\approx-\frac{1}{2}m^2\alpha^2\theta'+V_0$. Using Eq.~(\ref{eq:dI}), the canonical and non-canonical cases should be effectively described by $V_{\rm C}(I)=-I+V_0$ and $V_{\rm NC}(I)=-I^{-\frac{2}{3}}+V_0$, respectively. For the canonical case, we have the same prediction as the non-canonical $n=1$ case discussed above with $\tilde{r}=0.1067, \, N_e=37.5$. The numerical results for the non-canonical case are presented below.

With $(m, \, \lambda, \, \Lambda^4, \, f)_{\rm C}=(0.457, \, 500 \, m^2, \, 1\times 10^{-6}\, m^2, \, 0.001)$ and $(m, \, \lambda, \, \Lambda^4, \, f)_{\rm NC}=(0.0677, \, 150 \, m^2, \, 5\times 10^{-7}\, m^2, \, 0.001)$, we find $(\tilde{r}, \, n_s, \, n_r, \, \Delta_R^2, \, N_e)_{\rm C}=(0.16, \, 0.96, \, -8\times 10^{-4}, \, 2.2\times 10^{-9}, \, 49.5)$ and $(\tilde{r}, \, n_s, \, n_r, \, \Delta_R^2, \, N_e)_{\rm NC}=(0.1611, \, 0.96, \, -7.97\times 10^{-4}, \, 2.2\times 10^{-9}, \, 49.61)$. It is noticable that the numerical results of all three models with $W(r)=-\frac{1}{2}m^2r^2+\frac{\lambda}{4} r^4+\frac{m^4}{4\lambda}$ considered in this paper coincide with the prediction of a $V(I) \sim I^2$ model. This may be a result of the inflationary process occuring close to the minimum of the potential where the potential can be described as $\sim I^2$.

\section{Conclusions}
\label{sec:Conclusions}
We have analyzed and compared a variety of two-field inflation models with one or two axions, in particular Dante's inferno/waterfall-type models and spiral inflation models. These two classes of models are described by equivalent potentials, but differ  in the kinetic terms for the  fields, or equivalently the field-space metric. We have found that, not surprisingly, the field-space metric plays an important role in predictions for inflationary observables, with spiral inflation models generally predicting a smaller number of $e$-folds $N_e$ and tensor-to-scalar ratio $\tilde{r}$ than the Dante's inferno model with the same potential. Whereas the Dante's waterfall scenario yields a phenomenologically viable parameter space,  the corresponding spiral inflation model appears to face tighter phenomenological constraints. 

In some of the recent spiral inflation literature, observables were calculated using a mass-matrix formalism rather than appealing to a single-field effective description. It has been suggested that the single-field description, which maps these models to chaotic-inflation type models during inflation, is not generally valid  \cite{Barenboim:2015lla}. We have argued that a single-field description which maps these models into chaotic inflation models is valid (until the end of inflation, at which point the multi-field nature of the models is indeed important), and we constructed the  mass matrix relevant for comparison with the single-field description. The geometric approach taken here can be generalized to other multi-field models, but is simplified in spiral-inflation models by their nearly circular field-space trajectories. 

The single-field description  relates  observables in Dante's inferno-type models to those in spiral inflation models with related potentials, which is a type of duality between inflation models. Finally, we note that both the Dante's inferno and spiral inflation models have a flat field space, albeit in different parametrizations. It would be worthwhile to classify the effects of field-space curvature on inflation models with potential trenches, generalizing the models analyzed here.

\begin{acknowledgments}  
We thank Jack Donahue and Anuraag Sensharma for collaboration at early stages of this work. 
We also thank Gabriela Barenboim, Chris  Carone and Wan-Il Park for useful conversations.
This work was supported by the NSF under Grant PHY-1519644. 
J.O. thanks the William \& Mary Research
Experiences for Undergraduates (REU) program for its support via NSF Grant PHY-1359364 and the DeWilde Summer Research 
Fellowship of the College of William \& Mary. 
\end{acknowledgments}

\end{document}